\documentclass[]{spie}  

\usepackage{amsmath}
\usepackage{amssymb}
\usepackage{amsxtra}
\usepackage{latexsym}
\usepackage{cite}
\usepackage{booktabs}
\usepackage[pdfpagemode=UseNone,pdfstartview=FitH,pdfview=FitH,colorlinks=true,
 pdftitle=Perceptually-Driven\ Video\ Coding\ with\ the\ Daala\ Video Codec,
 pdfauthor=Xiph.Org]{hyperref}
\usepackage{color}
\usepackage{url}
\usepackage{caption}
\usepackage{subcaption}
\usepackage{multirow}
\usepackage[pdftex]{graphicx}

\newif\ifpdf
\ifx\pdfoutput\undefined
\pdffalse
\else
\pdfoutput=1
\pdftrue
\fi
\ifpdf
\else

\fi

\title{Perceptually-Driven Video Coding \\ with the Daala Video Codec}

\author[a,b]{Yushin Cho}
\author[a,b]{Thomas J. Daede}
\author[a,b]{Nathan E. Egge}
\author[a,b]{Guillaume Martres}
\author[a,b]{Tristan Matthews}
\author[a,b]{Christopher Montgomery}
\author[a,b]{Timothy B. Terriberry}
\author[a,b]{Jean-Marc Valin}
\affil[a]{Xiph.Org Foundation, 21 College Hill Road, Somerville, MA 02144, USA}
\affil[b]{Mozilla Corporation, 331 E. Evelyn Ave., Mountain View, CA 94041, USA}

\authorinfo{Copyright 2016 Mozilla Foundation.
This work is licensed under
 \href{https://creativecommons.org/license/by/4.0/}{CC-BY 4.0}.\\
Send correspondence to \href{mailto:tterribe@xiph.org}{tterribe@xiph.org}.}

\begin{document} 
\maketitle

\begin{abstract}
The Daala project is a royalty-free video codec that attempts to compete with
 the best patent-encumbered codecs.
Part of our strategy is to replace core tools of traditional video codecs with
 alternative approaches, many of them designed to take perceptual aspects into
 account, rather than optimizing for simple metrics like PSNR.
This paper documents some of our experiences with these tools, which ones
 worked and which did not.
We evaluate which tools are easy to integrate into a more traditional codec
 design, and show results in the context of the codec being developed by the
 Alliance for Open Media.
\end{abstract}


\section{INTRODUCTION}
\label{sec:introduction}

Designing a royalty-free video codec is a challenging problem~\cite{Rea10}.
Using tools and techniques that differ substantially from those used by the
 common commercially-licensed codecs can reduce that challenge.
Over the last several years in the Daala project~\cite{daala-website}, we have
 experimented with a number of such tools.
Some techniques were long known by other researchers~\cite{WS91,SM98,Tra01},
 but not commonly used.
However, we thought they might not have been sufficiently developed at the time
 to be competitive with the usual approaches, or might be worth re-evaluating
 in a modern context due to changes in engineering trade-offs.
Other techniques we developed ourselves~\cite{Ter15a,VT15,EV15}, either to
 solve existing problems in new ways or to solve new problems raised by the
 other tools we chose.

Some of these techniques have been successful, while others did not pan out as
 well as we hoped.
As effort is now underway at the Alliance for Open Media (AOM) to create a
 royalty-free video codec based on the technologies of many industry partners,
 including Daala, it is worth analyzing these techniques in that context.
The Alliance is starting development of its first codec, AV1, from
 VP9~\cite{vp9-spec}, a more traditional codec design, and also has access to
 patent rights that we did not as part of the Daala project.
However, we think that some of the technology we created and many of the
 lessons we learned can usefully contribute to this effort.

All of the code from the various experiments presented below is available in
 either the Daala git repository~\cite{daala-git} or the AOM git
 repository~\cite{aom-git}, except where, as noted, it was
 done in an author's personal (but still public) repository.
All metrics results for Daala are taken from our Are We Compressed Yet (AWCY)
 website, using the \texttt{ntt-short-1} sequences and ``Entire Curve (old
 method)" approach for computing the average change in rate at equivalent
 quality~\cite{awcy-website} (except as noted).
It makes use of four objective metrics, whose implementations are available in
 the \texttt{tools} directory of the Daala git repository, which are PSNR,
 SSIM~\cite{WVSS04}, PSNR-HVS-M~\cite{PSECAL07}, and multiscale
 FastSSIM~\cite{CB11}.
Metric results for AV1 are also taken from AWCY, but using the newer
 \texttt{objective-1-fast} test set and the ``Report (Overlap)" approach for
 computing rate changes, and use regular MS-SSIM~\cite{WSB03}.
This is the methodology currently proposed in the testing draft of the Internet
 Engineering Task Force (IETF)'s NETVC working group~\cite{DNB16}.
The source code for Are We Compressed Yet is also
 available~\cite{awcy-git,rd-tool-git}.

\section{AN ANALYSIS OF DAALA CODING TOOLS}

While Daala uses the traditional hybrid video coding model of
 motion-compensated prediction followed by block-based transform coding, it
 differs in most of the details.
It uses \textit{lapped transforms}~\cite{Egg15} instead of an adaptive loop
 filter for deblocking, with intra prediction performed in the
 \textit{frequency domain} (see Section~\ref{sec:freq-domain-intra}) and with
 variable-block-sized \textit{Overlapped Block Motion Compensation} (OBMC) to
 eliminate blocking artifacts from the prediction stage~\cite{Ter15a}.
It quantizes transform coefficients with a technique we term \textit{Perceptual
 Vector Quantization}~\cite{VT15} and codes them with a \textit{non-binary}
 multiply-free arithmetic coder~\cite{Ter15b} based on the work of Stuiver and
 Moffat~\cite{SM98}.
An analysis on the current performance, challenges, and future potential of
 each of the major tools follows.

\subsection{Lapped Transforms}
\label{sec:lapped-transforms}

Viewed in isolation, lapped transforms are extremely effective.
They provide better energy compaction than the DCT with less computational cost
 than wavelets with similar energy compaction~\cite{Tra01}, while structurally
 eliminating blocking artifacts (one of the most annoying visual artifacts at
 low rates) at the cost of some increased ringing.
Virtually every transform-based audio codec since MP3 has been built on top of
 them, yet the only major video codec standard to use them has been VC-1, and
 then only in intra frames.

Unfortunately, a transform is just one part of a video codec.
Conceptually, lapped transforms are a very simple change to the standard
 block-based design: the adaptive loop filter is replaced by an invertible,
 non-adaptive post-filter, and the encoder runs the inverse as a pre-filter
 before a normal DCT and quantization stage.
However, in practice they introduce structural complications elsewhere in the
 design.
\begin{enumerate}
\item They cannot be used with standard spatial intra prediction, because the
 final pixel values of neighboring blocks are not available until after the
 post-filter is applied.
See Section~\ref{sec:freq-domain-intra} for how we attempt to address this.
\item They make the inter/intra mode decision problem more complicated.
See Section~\ref{sec:obmc} for a discussion of this issue.
\item As originally proposed~\cite{DLT05}, they introduce 2D dependencies into
 the transform block size decision, which makes the problem NP-hard.
\item They are less localized than a DCT, which makes localized changes in
 inter frames more difficult to code.
This is not an issue for still images, because intra prediction is generally
 less effective, and the content which must be coded by the transform stage is
 less localized.
\end{enumerate}

In theory, many of these are also true for an adaptive loop filter.
The final pixel values for a neighboring block are not available until after
 it runs, the size of the filter and the order it is applied may depend on the
 choice of a neighbor's block size (causing the same 2D dependency problems),
 and when prediction errors are present on only one side of a block edge, they
 can introduce spurious changes into pixels on the other side.
However, a typical encoder ignores all of these effects, trusting that it can
 simply optimize the output of the transform and quantization stage.
This assumes that because the loop filter is adaptive, it will not make the
 result worse (or not enough to matter).
With lapped transforms, the post-filter is not adaptive.
It gets applied even when there is little to no horizontal or vertical
 correlation across a transform block edge.
That makes these issues impossible to ignore.

\subsubsection{Filter Ordering and Block Size Decision}
\label{sec:block-size-decision}

\begin{figure}[htb]
\center
\begin{subfigure}[c]{0.25\textwidth}
\center
\includegraphics[width=0.8\textwidth]{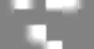}
\caption{DC basis functions.}
\label{fig:basis-weirdness}
\end{subfigure}%
\begin{subfigure}[c]{0.2222\textwidth}
\center
\includegraphics[height=0.75in]{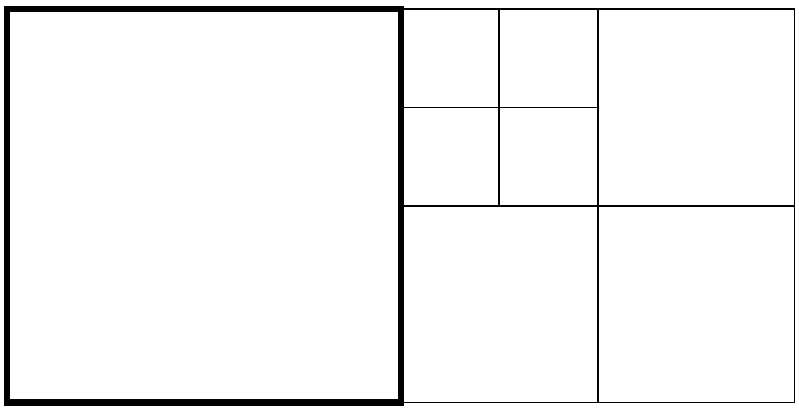}
\caption{Example block sizes.}
\label{fig:lapping-example}
\end{subfigure}%
\begin{subfigure}[c]{0.3056\textwidth}
\center
\includegraphics[height=0.75in]{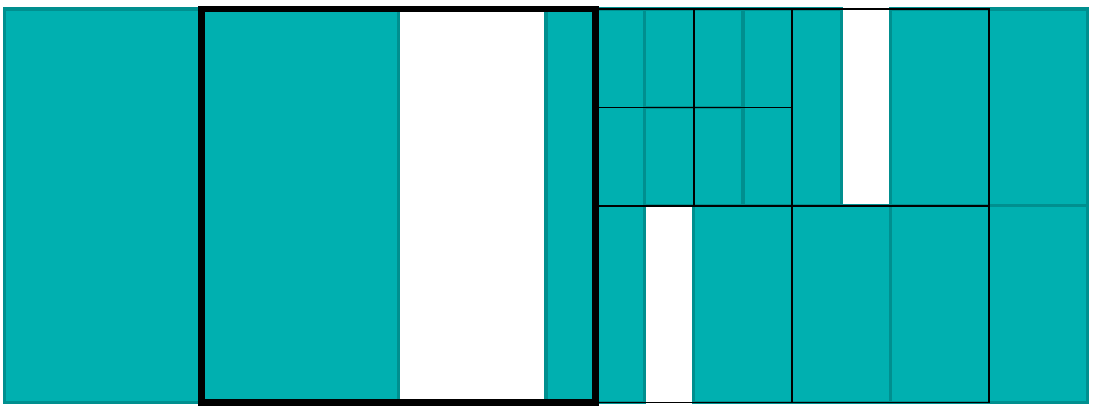}
\caption{Lapping vertical edges.}
\label{fig:lapping-example-vert}
\end{subfigure}%
\begin{subfigure}[c]{0.22222\textwidth}
\center
\includegraphics[height=1.5in]{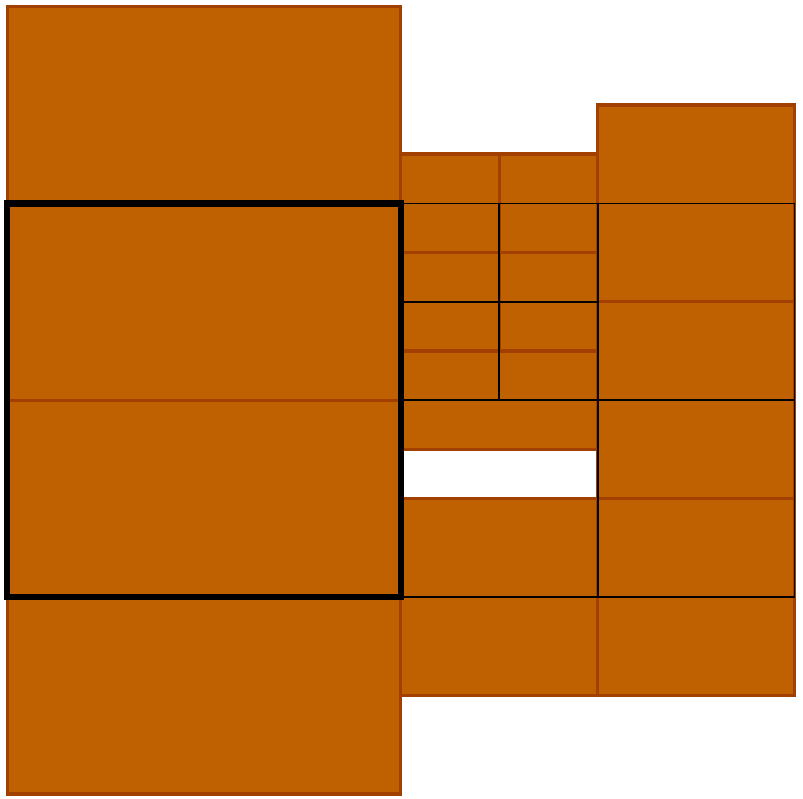}
\caption{Horizontal edges.}
\label{fig:lapping-example-horiz}
\end{subfigure}
\caption{Example DC basis functions for randomly chosen transform block
 sizes~(\ref{sub@fig:basis-weirdness}) and an (unrelated) example block size
 partitioning along with the vertical and horizontal lapping applied to each
 edge with the full lapping
 scheme~(\ref{sub@fig:lapping-example}-\ref{sub@fig:lapping-example-horiz}).}
\label{fig:lapping-examples}
\end{figure}

The 2D dependency problem is perhaps the most severe.
In our original design, we apply the post-filter in the same order as a
 traditional adaptive loop filter.
We apply the post-filter to each row of pixels that crosses a vertical block
 edge first (see Figure~\ref{fig:lapping-example-vert}), followed by each column
 of pixels that crosses a horizontal block edge (see
 Figure~\ref{fig:lapping-example-horiz}).
This maximizes parallelism (all row filters can be applied in parallel,
 followed by all column filters) and minimizes the buffering required.
The pre-filter is applied in the opposite order.
The size of the pre-/post-filter for each edge is the maximum that does not
 penetrate more than halfway into any neighboring transform block.
That is, we use the largest overlap we can apply that will not introduce
 any additional dependencies on the order the filters are applied, under the
 constraint that the overlap size is constant for a whole transform block
 edge.
E.g., a $16\times 16$ transform block next to an $8\times 8$ transform block
 and two $4\times 4$ transform blocks uses a $4$-tap filter along the
 entire shared edge (see Figure~\ref{fig:lapping-example-vert}).
We term this scheme the ``full lapping'' scheme.

This scheme has two issues.
The first is that the filtering order can produce some very strangely shaped
 basis functions.
To demonstrate this, we generated a frame with randomly chosen block sizes and
 set the DC coefficient in some blocks to a large, positive value.
The result can be seen in Figure~\ref{fig:basis-weirdness}.
Although disconcerting, we never observed visual artifacts in real images with
 similar patterns, so this problem is mostly theoretical (and one that affects
 adaptive loop filters, as well).

The second problem is the much more practical issue that we have no idea how to
 search for the optimal transform block size.
In order to compute the rate and distortion required for a traditional
 rate-distortion (R-D) search, we needed to know what filters were applied
 along each edge, which requires knowing the size of neighboring transform
 blocks.
It is not even enough to know the size of the lapping on each edge.
If a neighbor on the other side of a vertical edge is split, then the
 pre-filter is applied along the horizontal edge between them before it is
 applied along the vertical edge in the current block, changing the filter
 output.
That is, in the example of Figure~\ref{fig:lapping-example}, the pre-filters in
 Figure~\ref{fig:lapping-example-horiz} are applied before the pre-filters in
 Figure~\ref{fig:lapping-example-vert}.
Therefore, it is necessary to know \textit{which} neighboring block is
 $8\times 8$ and which are $4\times 4$ to properly determine the output of the
 pre-filter in the $16\times 16$ block.
The corresponding dependency graph does not have a tree structure which admits
 a simple dynamic programming solution.

Our initial solution to this was to use a heuristic to select block sizes
 up-front, based on a perceptual estimate of the visibility of ringing
 artifacts.
While this seemed to do vaguely reasonable things for intra frames,
 we did not have a known-optimal algorithm to compare it to, so we could not be
 certain how far off it really was.
We \textit{were} able to determine it was doing a poor job for inter frames:
 simply biasing the decision away from $4\times 4$ and $8\times 8$ blocks for
 large quantizers reduces the bitrate by more than $5\%$ on all of the metrics
 we tested\footnote{See commit be43a7c18ea7 in the Daala git
 repository~\cite{daala-git}}.

\subsubsection{Fixed Lapping}
\label{sec:fixed-lapping}

In order to permit a true RDO search, we change two things~\cite{DEVMT16}.

First, we change the lapping order.
Dai et al. propose filtering on $16\times 16$ macroblock (MB) boundaries first,
 followed by interior block edges (using the same order as the full lapping
 scheme in the interior)~\cite{DLT05}.
During block size decision, they use the smallest possible pre-filter on MB
 boundaries, and then exhaustively search partitions of the interior using
 the maximum possible overlap on internal block boundaries (as we described
 above).
After block size decision, in a second pass, they filter MB boundaries with
 the maximum possible overlap for actual encoding, as in our full lapping
 scheme.
In their implementation, MBs can be split down to $4\times 4$, producing a
 total of $17$ possible quad-tree subdivisions, which is still tractable to
 search exhaustively.
However, at the time we conducted these experiments, we already supported
 block sizes up to $32\times 32$, which would require searching $83,522$
 partitions, and would later extend this to $64\times 64$ (more than $48$
 quintillion partitions).

Instead, we apply what they do on MB boundaries recursively.
Starting at the largest block size, what we term a superblock (SB), we
 pre-filter the exterior SB edges first.
If the block is split, then we pre-filter the interior edges, and then recurse
 into each quadrant.
Again, when post-filtering, we reverse the order.

This reduces the amount of potential parallelism, but still allows filtering
 all SB interiors in parallel, followed by all vertical SB edges in parallel
 and then all horizontal SB edges.
But this change alone already makes our problem simpler.
Returning to the example of Figure~\ref{fig:lapping-example}, we no longer need
 to know which adjacent blocks are $4\times 4$, since the edge between them
 will always be pre-filtered after the edge they share with the $16\times 16$
 block.
All that remains is the issue of the filter size.

\begin{figure}[htb]
\center
\begin{subfigure}[c]{0.25\textwidth}
\center
\includegraphics[height=1.25in]{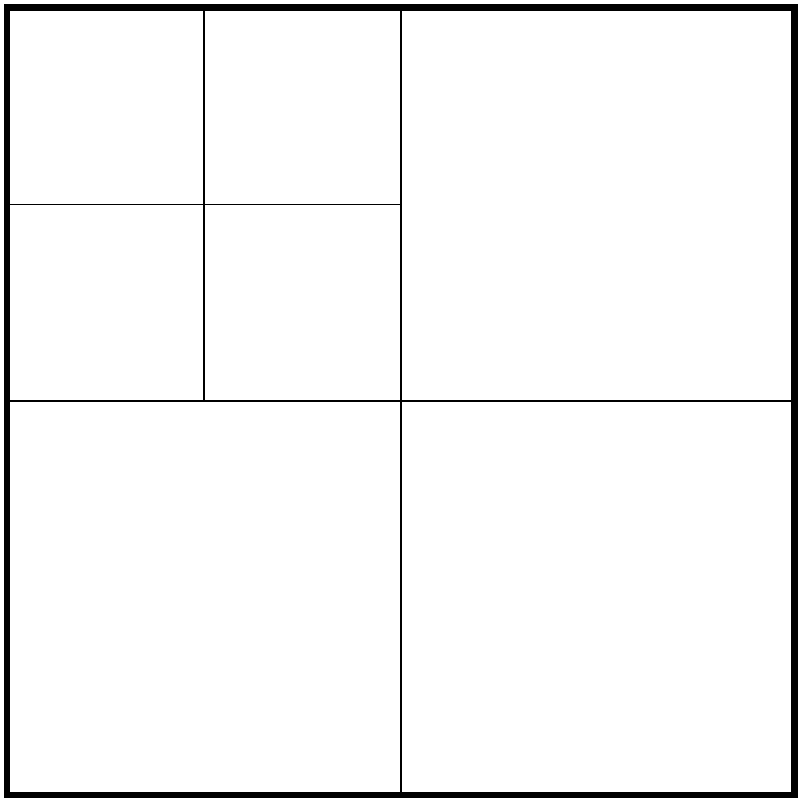}
\caption{Example block sizes.}
\label{fig:fixed-lapping-example}
\end{subfigure}%
\begin{subfigure}[c]{0.25\textwidth}
\center
\includegraphics[height=1.25in]{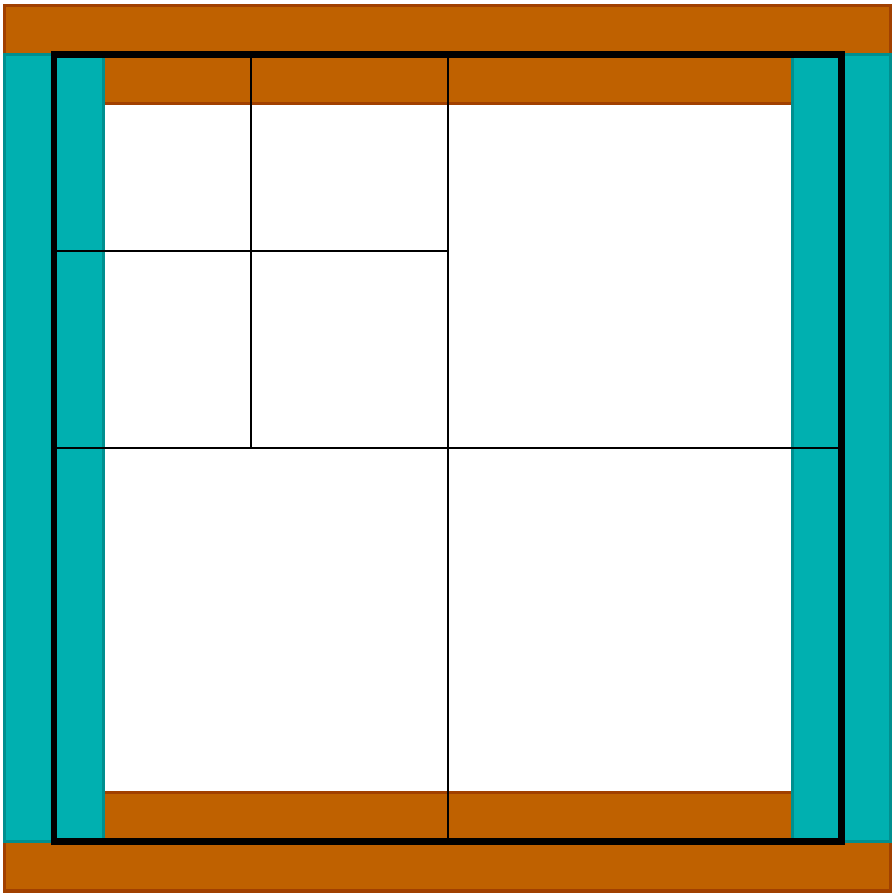}
\caption{Exterior edges.}
\label{fig:fixed-lapping-exterior}
\end{subfigure}%
\begin{subfigure}[c]{0.25\textwidth}
\center
\includegraphics[height=1.25in]{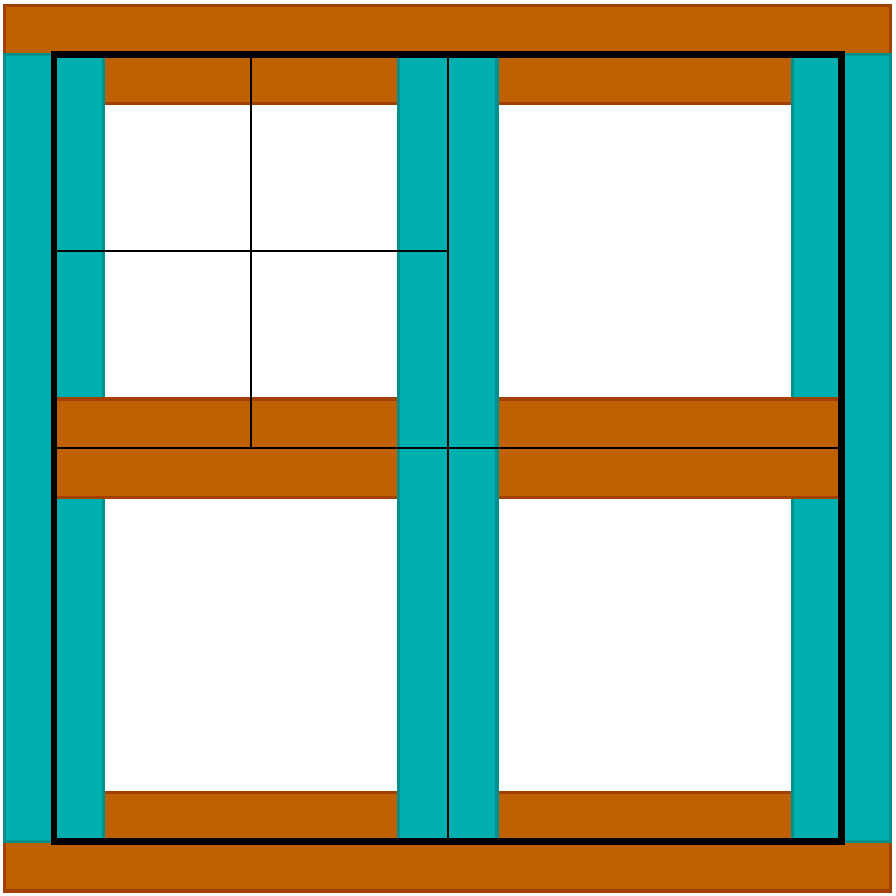}
\caption{Interior edges.}
\label{fig:fixed-lapping-int1}
\end{subfigure}%
\begin{subfigure}[c]{0.25\textwidth}
\center
\includegraphics[height=1.25in]{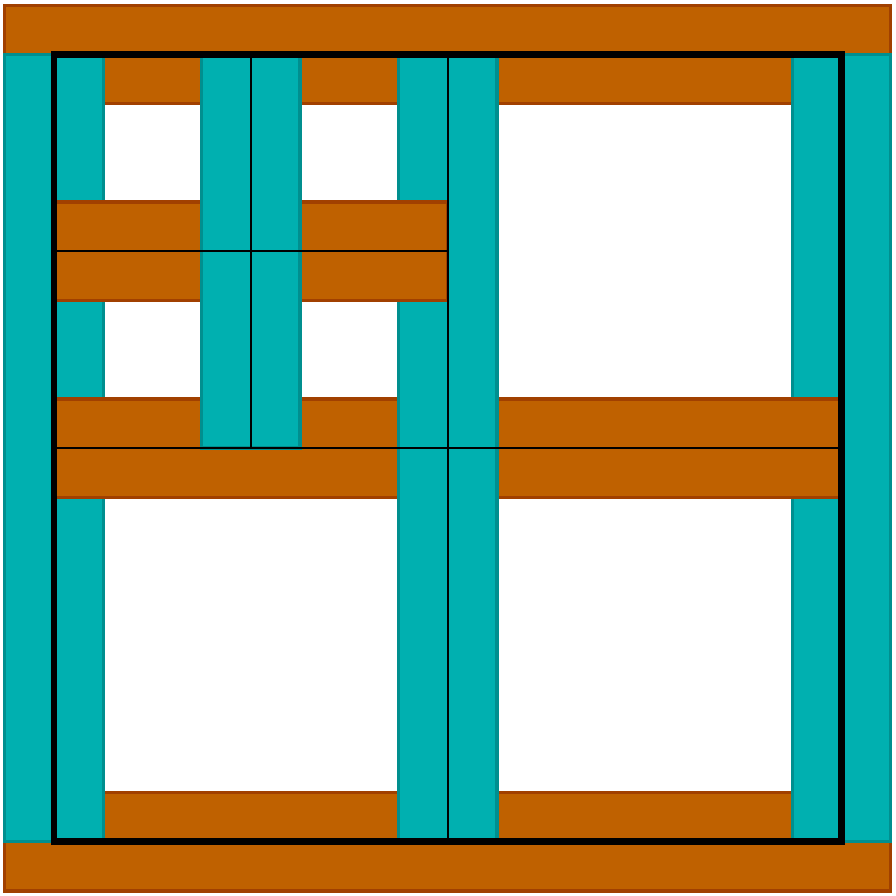}
\caption{Recursion.}
\label{fig:fixed-lapping-int2}
\end{subfigure}
\caption{Example demonstrating the pre-filtering order for the fixed lapping
 scheme.
At each step, the horizontal edges are pre-filtered before the vertical ones,
 to minimize buffering when applying the post-filter.}
\label{fig:fixed-lapping}
\end{figure}

To solve this problem, we make our second change: we make the filter size
 constant.
At first we used $8$-point filters until we had split blocks down to
 $4\times 4$, and then used $4$-point filters on the final interior edges.
However, currently we use $4$-point lapping everywhere, as it provides
 substantial reductions in ringing, at the cost of some blocking artifacts and
 some detail preservation.
We term this the ``fixed lapping scheme'', illustrated in
 Figure~\ref{fig:fixed-lapping}.

With those two changes, it is now possible to compute R-D optimal transform
 block sizes.
After pre-filtering the exterior edges, we try splitting the block and
 pre-filtering the interior edges.
Then we compute the optimal partitioning of each of the four quadrants
 (recursively), then apply the post-filter to the interior edges.
At this point, although we have not post-filtered the exterior edges, it is
 still possible to do an apples-to-apples comparison of the distortion between
 the split and no-split cases.

The result is large rate reductions on all metrics: $10.3\%$ and $12.3\%$ on
 PSNR and SSIM, respectively, and $4.5\%$ and $5.2\%$ on PSNR-HVS and Fast
 MS-SSIM, respectively, compared to the full lapping scheme\footnote{With the
 initial $8$-point lapping.
See commits 34b6d5950541 through 1a606b2c8e63 in the Daala git
 repository~\cite{daala-git}.
This also included the addition of a bilinear smoothing filter on intra
 frames~\cite{DEVMT16}, to reduce the blocking artifacts caused by the
 reduction in lapping.
Despite visual reductions in blocking, it actually slightly reduced metrics,
 and was later removed after it was found that $64\times 64$ DCTs and our
 deringing filter provided the same benefits, and it was no longer helping on
 most images.
See commit eca1b5a207bf.}.
For intra frames, the results showed similarly large gains, indicating that
 even though our heuristic looked ``vaguely reasonable" in this case, it was
 not actually doing a good job.
We also experimented with an approach similar to Dai et al.'s, where block size
 decisions are made using the fixed lapping size, but the actual video is
 encoded with the full lapping sizes (though still applied in recursive order).
When optimizing for PSNR (using flat quantization and no activity masking),
 this also achieves large gains: a $9.2\%$ reduction in bitrate\footnote{See
 the \texttt{var\_lap\_jm\_exp\_psnr2} branch of Yushin Cho's personal Daala
 repository: \url{https://git.xiph.org/users/yushin/daala.git}.}.
This was an encoder-only change, and we verified the bitstreams were decodable
 by the mainline decoder at the time.
This shows that most of the improvements of the fixed-lapping approach simply
 came from making better decisions than our previous heuristics.
However, we opted to keep the fixed lapping size.
The gains with the full lapping size are not as large as when using fixed
 lapping, and it is not obvious how to tune such an encoder, since the actual
 data that gets encoded can be significantly different than what is used to
 make decisions.

It might also be possible to explicitly signal the size of the pre-filters to
 use.
The search remains tractable in the interior of a superblock, but some kind of
 heuristic would be required for the lapping on edges between superblocks, as
 they retain the original 2D dependency problem.
Given that most of the motivation for larger lapping is the reduction in
 blocking artifacts in large regions that are either smooth or uniformly
 textured (where the largest transform blocks are used), this is a problem that
 would need to be solved.
This is an area we have not explored.

\subsubsection{Lapping vs. Loop Filters}

The same recursive filter application order can also be applied to a standard
 adaptive loop filter.
This allows an encoder to consider the effect of the loop filter when
 making transform block size decisions.
We modified Daala to use the adaptive loop filter from the Thor video
 codec~\cite{FBMDZ16} in place of lapping through a compile-time option.
When enabled, this produces bitrate reductions between $0.4\%$ and $2.6\%$ on
 video, depending on the metric used\footnote{See commit ae4a21325d62 in the
 Daala git repository~\cite{daala-git}.}.
However, when used on still images it produces bitrate increases between
 $2.5\%$ and $3.2\%$ (with the exception PSNR-HVS, where bitrate falls by
 $0.3\%$)\footnote{Using the same commit.
Tested on \texttt{subset1}, a set of 50 images sampled from Wikipedia and
 downsampled to about one megapixel on AWCY.}.
The visual impact is more difficult to judge.
We can also get metrics gains on video by disabling lapping completely, despite
 the obvious blocking and loss of detail preservation that causes.
However, we have not attempted to do a formal subjective comparison between
 lapped transforms and adaptive loop filters, and the results would probably be
 too close to justify the time and effort.
So, although we believe we have made good strides in understanding how to use
 lapping in a modern video codec, it is not obviously technically superior.

In the context of the Alliance for Open Media, using lapping would require so
 many structural changes to the codec that it would not be possible to do and
 still meet our time-to-market goals, so it is not currently being pursued.
However, the recursive filter ordering may still provide some benefit in terms
 of improved encoder decisions and potentially eliminating artifacts along
 the lines of Figure~\ref{fig:basis-weirdness}.
This is something we hope to experiment with in the future.

\subsection{Frequency-Domain Intra Prediction}
\label{sec:freq-domain-intra}

There are ample reasons in the research literature to suggest that the issue of
 intra prediction with lapped transforms raised in
 Section~\ref{sec:lapped-transforms} is a solvable problem.
For example, de Oliveira and Pesquet-Popescu show gains on five images by
 integrating lapped transforms into a version of H.264/AVC restricted to
 $8\times 8$ blocks and performing intra prediction from four pixels farther
 away~\cite{OP11}.
This is a promising result, but raises several practical issues.
First, they restrict the prediction to four of nine intra modes in the lapped
 transform case, because they do not have sufficient pixels available for the
 other modes.
Second, determining which pixels are available for prediction becomes much more
 complicated when introducing multiple block sizes, especially with the full
 lapping scheme (which we were still using when performing most of our intra
 experiments).
This approach also has a significant impact on the potential parallelism, since
 enough post-filtering must be applied to make the required pixels available
 before the current block can be reconstructed (and then post-filtered), but
 this is a problem with spatial intra prediction generally.

We draw inspiration from Xu et al., who recharacterize traditional
 spatial-domain intra prediction as an ``intra-predictive
 transform''~\cite{XWZ09}:
\begin{align}
  \begin{bmatrix} \mathbf{\tilde{x}} \\ \mathbf{y} \end{bmatrix} & =
   \begin{bmatrix} \mathbf{I} & \mathbf{0} \\
   \mathbf{0} & \mathbf{DCT} \end{bmatrix}
   \begin{bmatrix} \mathbf{I} & \mathbf{0} \\
   -\mathbf{E} & \mathbf{I} \end{bmatrix}
   \begin{bmatrix} \mathbf{\tilde{x}} \\ \mathbf{x} \end{bmatrix}\ .
\end{align}
Here, $\mathbf{\tilde{x}}$ represents the pixels used as predictors, which are
 unchanged by the transform, and $\mathbf{E}$ represents the prediction
 process.
In the simplest case, where $\mathbf{\tilde{x}}$ is a single pixel,
 $\mathbf{E}$ might be a column of $1$s.
Then $\mathbf{y}$ would be the DCT of the residual after extending
 $\mathbf{\tilde{x}}$ into the block and subtracting it from each pixel in
 $\mathbf{x}$.

This approach can be extended to move the prediction into the frequency domain.
We can apply any invertible transform to $\mathbf{\tilde{x}}$ (for example, the
 DCT) and still achieve identical output for $\mathbf{y}$ by applying the
 inverse transform to $\mathbf{E}$.
That is,
\begin{align}
  \begin{bmatrix} \mathbf{\tilde{y}} \\ \mathbf{y} \end{bmatrix} & =
   \begin{bmatrix} \mathbf{I} & \mathbf{0} \\
   \mathbf{0} & \mathbf{DCT} \end{bmatrix}
   \begin{bmatrix} \mathbf{I} & \mathbf{0} \\
   -\mathbf{E}\cdot\mathbf{DCT}^{-1} & \mathbf{I} \end{bmatrix}
   \begin{bmatrix} \mathbf{DCT} & \mathbf{0} \\
   \mathbf{0} & \mathbf{I} \end{bmatrix}
   \begin{bmatrix} \mathbf{\tilde{x}} \\ \mathbf{x} \end{bmatrix}\ .
\end{align}
Multiplying together the first two matrices on the right hand side produces
\begin{align}
  \begin{bmatrix} \mathbf{\tilde{y}} \\ \mathbf{y} \end{bmatrix} & =
   \begin{bmatrix} \mathbf{I} & \mathbf{0} \\
   -\mathbf{DCT}\cdot\mathbf{E}\cdot\mathbf{DCT}^{-1} & \mathbf{DCT} \end{bmatrix}
   \begin{bmatrix} \mathbf{DCT} & \mathbf{0} \\
   \mathbf{0} & \mathbf{I} \end{bmatrix}
   \begin{bmatrix} \mathbf{\tilde{x}} \\ \mathbf{x} \end{bmatrix}\ .
\end{align}
Then, observing that we can also pre-transform $\mathbf{x}$ if we correct for
 it in the next step we have
\begin{align}
  \begin{bmatrix} \mathbf{\tilde{y}} \\ \mathbf{y} \end{bmatrix} & =
   \begin{bmatrix} \mathbf{I} & \mathbf{0} \\
   -\mathbf{DCT}\cdot\mathbf{E}\cdot\mathbf{DCT}^{-1} & \mathbf{DCT} \end{bmatrix}
   \begin{bmatrix} \mathbf{I} & \mathbf{0} \\
   \mathbf{0} & \mathbf{DCT}^{-1} \end{bmatrix}
   \begin{bmatrix} \mathbf{I} & \mathbf{0} \\
   \mathbf{0} & \mathbf{DCT} \end{bmatrix}
   \begin{bmatrix} \mathbf{DCT} & \mathbf{0} \\
   \mathbf{0} & \mathbf{I} \end{bmatrix}
   \begin{bmatrix} \mathbf{\tilde{x}} \\ \mathbf{x} \end{bmatrix}\ ,
\end{align}
 which simplifies to
\begin{align}
  \begin{bmatrix} \mathbf{\tilde{y}} \\ \mathbf{y} \end{bmatrix} & =
   \begin{bmatrix} \mathbf{I} & \mathbf{0} \\
   -\mathbf{DCT}\cdot\mathbf{E}\cdot\mathbf{DCT}^{-1} & \mathbf{I} \end{bmatrix}
   \begin{bmatrix} \mathbf{DCT} & \mathbf{0} \\
   \mathbf{0} & \mathbf{DCT} \end{bmatrix}
   \begin{bmatrix} \mathbf{\tilde{x}} \\ \mathbf{x} \end{bmatrix}\ ,
\end{align}
Letting the matrix
 $\mathbf{F}=\mathbf{DCT}\cdot\mathbf{E}\cdot\mathbf{DCT}^{-1}$, we call
 $\mathbf{F}$ the \textit{frequency domain predictor} that is equivalent to
 $\mathbf{E}$.
For any spatial domain predictor $\mathbf{E}$, we can derive an equivalent
 frequency-domain predictor $\mathbf{F}$ by passing it through the
 corresponding transforms.

Since we know that lapped transforms have similar frequency-domain behavior to
 that of the DCT, generally, we posit that if we replace the DCT by a lapped
 transform, there exist similar frequency-domain predictors.
We can no longer claim direct equivalence to a spatial-domain predictor, and it
 is not at all clear that the resulting predictors admit efficient
 implementations.
Instead of attempting to design such predictors theoretically, we train them
 from real data, following an approach similar to that used for directional
 transforms by Sezer et al.~\cite{SHG08}

\begin{figure}[htb]
\center
\begin{subfigure}[b]{0.33\textwidth}
\center
\includegraphics[width=0.9\columnwidth]{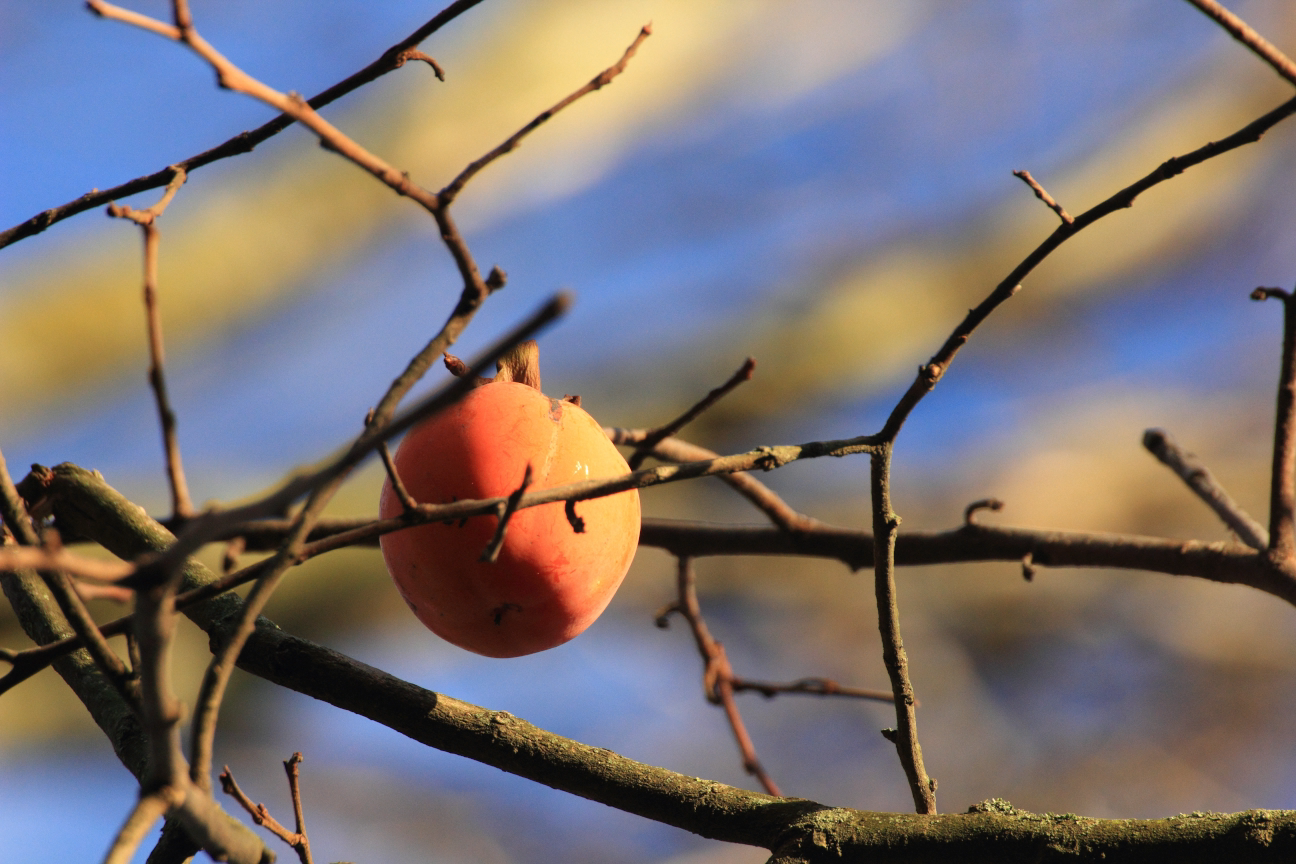}
\caption{Original image.}
\label{fig:intra-modes-orig}
\end{subfigure}%
\begin{subfigure}[b]{0.33\textwidth}
\center
\includegraphics[width=0.9\columnwidth]{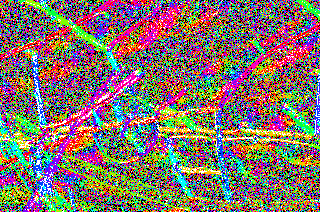}
\caption{VP8 intra classification.}
\label{fig:intra-modes-vp8}
\end{subfigure}%
\begin{subfigure}[b]{0.33\textwidth}
\center
\includegraphics[width=0.9\columnwidth]{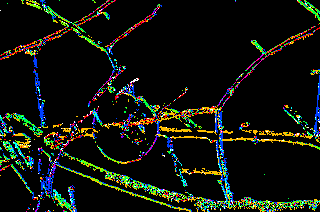}
\caption{After K-means iteration.}
\label{fig:intra-modes-iter}
\end{subfigure}
\caption{Classification of blocks by intra mode during training.
(\ref{sub@fig:intra-modes-orig}) shows the original image, followed by the
 classification of $4\times 4$ blocks by their best intra prediction mode
 (\ref{sub@fig:intra-modes-vp8}) using the VP8 predictors and
 (\ref{sub@fig:intra-modes-iter}) after several iterations of training.
Black represents the DC mode, white the True Motion mode, and the remaining
 colors the directional modes, arranged by hue.}
\label{fig:intra-modes}
\end{figure}

We classify blocks on a set of $1,000$ natural images sampled from Wikipedia
 (\texttt{subset3} on AWCY), each downsampled to about one megapixel, using the
 ten intra prediction modes from VP8~\cite{vp8-rfc}, extended to work with
 arbitrary block sizes, picking the one that minimizes the Sum of Absolute
 Transformed Difference (SATD).
Then, we train ten frequency domain predictors using least-squares regression
 on the blocks for each mode, and re-classify the blocks using the new
 predictors, and iterate this process, as illustrated in
 Figure~\ref{fig:intra-modes}.
Initially we weighted the blocks based on how well the best predictor fared
 compared to the second best, in order to reduce the impact of outliers, blocks
 which were either equally well predicted by several modes, or blocks not well
 predicted by any mode.
However, this did not appear to help, and we eventually disabled it.

\begin{figure}[htb]
\center
\begin{subfigure}[b]{0.34\textwidth}
\center
\includegraphics[width=0.9\columnwidth]{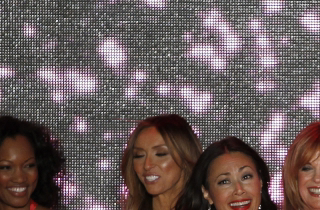}
\caption{Original image.}
\label{fig:end-of-show-orig}
\end{subfigure}%
\begin{subfigure}[b]{0.34\textwidth}
\center
\includegraphics[width=0.9\columnwidth]{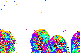}
\caption{Intra classification.}
\label{fig:end-of-show-intra}
\end{subfigure}%
\begin{subfigure}[b]{0.28\textwidth}
\center
\includegraphics[width=0.9\columnwidth]{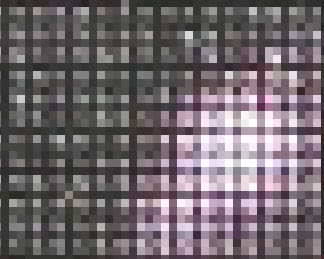}
\caption{Close-up of texture.}
\label{fig:end-of-show-closeup}
\end{subfigure}
\caption{An image with a patterned background texture.
 (\ref{sub@fig:end-of-show-orig}) A cropped region of the original image,
 (\ref{sub@fig:end-of-show-intra}) the corresponding classification by
 intra modes after training, and
 (\ref{sub@fig:end-of-show-closeup}) a close-up of the background texture
 pattern.
Intra mode colors are the same as in Figure~\ref{fig:intra-modes}.}
\label{fig:end-of-show}
\end{figure}

Beyond the initial classification using the VP8 intra predictors, we imposed no
 constraint to ensure that the predictors remained directional, though they
 generally did so\footnote{See
 \url{https://people.xiph.org/~xiphmont/demo/daala/demo2.shtml}
 for an interactive demonstration of the $4\times 4$ predictors.}.
Interestingly, one mode did diverge significantly from its original use: VP8's
 ``True Motion" mode.
This mode forms its prediction as $p_{ij}=U_j+L_i-UL$ where $(i,j)$ is the
 coordinate of the pixel being predicted, $U_j$ is the corresponding pixel on
 the edge of the neighboring block above, $L_i$ is the corresponding pixel on
 the edge of the neighboring block to the left, and $UL$ is the pixel at the
 corner of the neighbor above and to the left.
While in VP8 its primary utility is in predicting gradients, after training we
 found that it was used in highly textured regions, as illustrated in
 Figure~\ref{fig:end-of-show}.
Closer inspection suggests that what the mode is actually doing is copying the
 contents of its neighboring blocks at a half-cycle offset (that is, at an
 offset half the size of the block).
It seems possible to manually design such an intra mode, in either the
 frequency or the spatial domain, though we are not aware of anyone doing so.

\begin{figure}[htb]
\center
\begin{subfigure}[b]{0.4\textwidth}
\center
\includegraphics[width=\columnwidth]{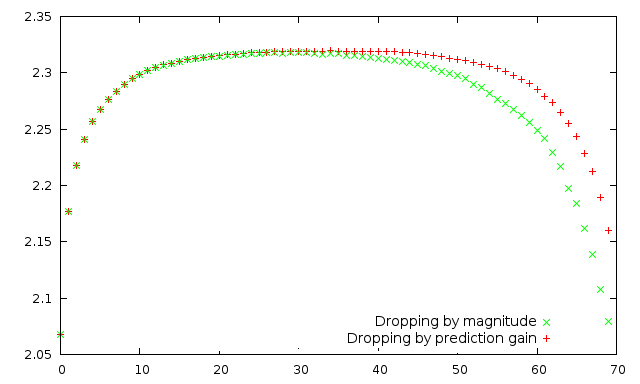}
\caption{Average prediction gain (dB).}
\end{subfigure}
\begin{subfigure}[b]{0.4\textwidth}
\center
\includegraphics[width=\columnwidth]{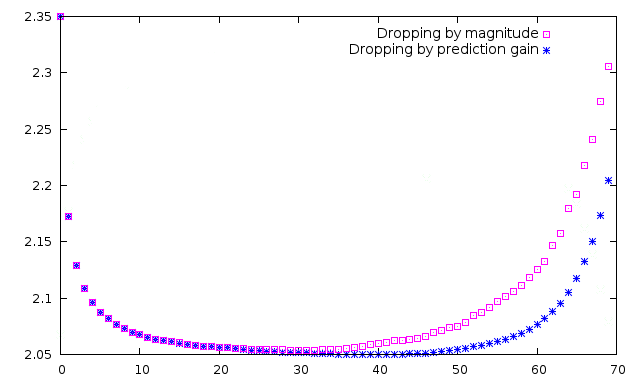}
\caption{Estimated lossless bitrate (bits per pixel).}
\end{subfigure}%
\caption{Comparison of coefficient zeroing strategies at each iteration ($x$
 axes) of predictor training for $4\times 4$ blocks on $50$ images.
Estimated bitrates are based on the entropy of coding Laplacian distributions
 with a known variance and do not include the cost to code the prediction
 modes.}
\label{fig:mag-vs-ref}
\end{figure}

In order to ensure that our predictors are computationally feasible, we set a
 complexity budget of four multiplies per output coefficient.
We force a fixed number of additional entries in the prediction matrix
 $\mathbf{F}$ to zero at each iteration until this budget is met.
We experimented with picking the smallest entries to force to zero or
 picking the ones that had the smallest impact on prediction gain (that is,
 that gave the smallest increase in the geometric mean of the variance of the
 prediction residuals), and the latter, though much slower to
 train, appears universally better (see Figure~\ref{fig:mag-vs-ref}).

We ran this procedure, with and without the sparsity constraint, on our $1,000$
 image training set using fixed $4\times 4$, $8\times 8$, and $16\times 16$
 transform block sizes, and compared the prediction gain to the spatial domain
 VP8 predictors with the DCT.
The results are summarized in Table~\ref{tab:fdip-pg-results}.
Coding gain and prediction gain are computed using actual measured variances of
 the full 2D transform coefficients, rather than the traditional first-order
 auto-regressive AR(1) model with correlation coefficient $\rho=0.95$.

\begin{table}[tb]
\begin{center}
\begin{tabular}{l@{\hspace{0.15in}}rrrc@{\hspace{0.25in}}rrrc@{\hspace{0.15in}}rr}\toprule
               & \multicolumn{3}{l}{DCT+VP8 Intra} & & \multicolumn{3}{l}{LT+FDIP (full)} & & \multicolumn{2}{l}{LT+FDIP (sparse)} \\\cmidrule{2-4}\cmidrule{6-8}\cmidrule{10-11}
Transform size & $C_g$     & $P_g$     & Total     & & $C_g$     & $P_g$     & Total      & & $P_g$     & Total     \\\midrule
$4\times 4$    & $13.8511$ & $2.9154$  & $16.7665$ & & $14.9600$ & $2.2103$  & $17.1703$  & & $2.0799$  & $17.0399$ \\
$8\times 8$    & $15.1202$ & $0.86940$ & $15.9894$ & & $15.6468$ & $0.72486$ & $16.3716$  & & $0.58957$ & $16.2364$ \\
$16\times 16$  & $15.5870$ & $0.19483$ & $15.7818$ & & $15.8721$ & $0.37593$ & $16.2480$  & & $0.09541$ & $15.9675$ \\
\bottomrule\end{tabular}
\end{center}
\caption{Coding gains and prediction gains of the DCT with the VP8 intra
 predictors compared to lapped transforms and frequency-domain intra prediction
 (FDIP) with (full) and without (sparse) the predictor sparsity constraint.
The ``Total" columns represent the coding gain of the corresponding transform
 applied to the prediction residual.
All numbers are in decibels (dB, higher is better).}
\label{tab:fdip-pg-results}
\end{table}

At the $4\times 4$ level, the sparse predictors work fairly well, with little
 loss relative to the non-sparse versions.
The frequency-domain predictors show less prediction gain than their spatial
 counterparts, but this is because some of the correlation with neighboring
 blocks has already been removed by the lapping, which is accounted for in the
 higher coding gains of the lapped transforms.
The prediction gain for the full $16\times 16$ predictor actually appears to
 be higher than that of the spatial predictor, but at this size each prediction
 matrix has $262,144$ free parameters.
With only $1,000$ one megapixel training images, and ten modes to train, this
 gives less than two $16\times 16$ input image blocks per free parameter, so
 this result is probably due to over-fitting.
Sparsification reduces this to only $1,024$ free parameters per prediction
 mode, but this also eliminates most of the prediction gain.

Overall, the prediction gain for sizes larger than $4\times 4$ is
 disappointing, and the sparsity constraint has a fairly significant impact on
 the prediction quality.
In practice, we found that the training would spend a large portion of its
 non-zero matrix entries predicting DC, as this had the biggest impact on the
 prediction gain.
This was unsatisfying, as there are less expensive ways to predict DC
 coefficients (see Section~\ref{sec:simpler-intra}).

\subsubsection{Variable Block Sizes}
\label{sec:variable-block-size-intra}

Extending this to variable block sizes creates even more complications.
All of the prediction matrices trained above assume that each block is the same
 size as its neighbors.
In practice, they may be different sizes.
Training separate prediction matrices for every potential combination of
 neighboring block sizes is clearly impractical.

To handle this, we borrow a tool from the Opus audio codec: Time-Frequency (TF)
 resolution switching~\cite{VMTV13}.
The basic idea is to apply a Walsh-Hadamard Transform (WHT) to very cheaply
 merge multiple blocks of frequency-domain coefficients into a single, larger
 block with higher frequency resolution, or conversely to split a large block
 of frequency-domain coefficients into smaller blocks with higher time
 resolution.
The conversion is grossly approximate, but can be ``good enough" for some
 purposes.

Since all of our coefficient blocks are two-dimensional (unlike audio), we can
 implement a particularly cheap integer version that is a has orthonormal
 scaling (i.e., is $L_2$-norm preserving) and exactly reversible, with just
 seven additions and one shift.
In order to implement the $2\times 2$ WHT 
\begin{align}
  \begin{bmatrix} A & B \\ C & D \end{bmatrix} & =
   \frac{1}{2}\begin{bmatrix} 1 & 1 \\ 1 & -1 \end{bmatrix}
   \begin{bmatrix} a & b \\ c & d \end{bmatrix}
   \begin{bmatrix} 1 & 1 \\ 1 & -1 \end{bmatrix}\ ,
\end{align}
 we compute
\begin{enumerate}
\item $e=a+c$, $f=d-b$,
\item $g=\tfrac{1}{2}(e-f)$,
\item $B=g-b$, $C=g-c$,
\item $A=e-B$, $D=f+C$.
\end{enumerate}
This is significantly cheaper than other approaches described in the
 literature, such as that of Fukuma et al. (8~additions, 2~shifts, and
 1~negation)~\cite{FOIK99} or that used by JPEG XR (10~additions, 1~shift, and
 2~negations)~\cite{TSSRM08}, and is, to our knowledge, a novel result.

With this tool in hand, we can handle variable block sizes several ways.
For example, for each transform block, we can apply TF merging or splitting
 until our neighbors are all the same size as the current block, and then use
 the predictors trained for that size.
This is what we implemented first.

However, there is no requirement that our predictors be trained with neighbors
 that are the same size as the block being predicted.
Instead, we can TF all of our neighbors down to $4\times 4$, and then train
 predictors for each block size from the TF'd $4\times 4$ data.
During training, we transform all blocks at a fixed block size, and then TF
 the transform coefficients in a block's neighbors from that size down to
 $4\times 4$.
We also tried using the block size heuristic from
 Section~\ref{sec:block-size-decision} to choose block sizes, limiting the
 training for each size to blocks that would be encoded at that size, but
 this seemed to have little impact.
Using TF'd $4\times 4$ neighbors has a number of advantages:
\begin{enumerate}
\item Much simpler logic, since everything is converted to the same size, and
 we don't have to retransform the same blocks into multiple different TF
 resolutions when using them to predict multiple neighbors of different sizes.
\item Reduced memory footprint, since we only have to buffer one row of
 $4\times 4$ coefficient values, rather than all coefficients for an entire row
 of superblocks.
\item Less overfitting during training, since at $16\times 16$ we only have
 $53,248$ free parameters per prediction mode (before sparsification) instead
 of $262,144$.
\end{enumerate}
The difference between the two approaches on metrics is small.
The TF'd $4\times 4$ approach gives a tiny gain on PSNR and loses between
 $0.4\%$ and $1.4\%$ on our other metrics\footnote{See commit 2b03b324f7d6 in
 the Daala git repository~\cite{daala-git}.
These tests were done using the \texttt{rd\_collect.sh} script in the
 \texttt{tools} directory of that repository, since they predated AWCY}, but
 considering its other advantages we thought those losses acceptable.

\subsubsection{A Simpler Approach}
\label{sec:simpler-intra}

Although we are able to build a practical codec with lapped transforms and
 variable block sizes based on the frequency-domain intra prediction scheme
 described in the previous sections, we cannot say that it truly works well.
The promising results of images coded at a fixed $4\times 4$ block size
 disappear when larger block sizes are allowed, and over-fitting and training
 time, already a problem with $16\times 16$ transform blocks, pose a serious
 challenge for $32\times 32$ and above.
The prediction does show gains of a few percent on metrics, but not enough to
 justify the computational complexity of the approach: a sparse matrix multiply
 with four multiplies per predicted coefficient, multiplied by the number of
 modes that have to be searched in the encoder, with a large table of trained
 prediction parameters, and a row buffer of $4\times 4$ TF'd coefficient values
 (four times the size of the equivalent buffer for a spatial predictor).
It also has the disturbing tendency to occasionally produce large, erroneous,
 isolated high-frequency coefficients (likely due to over-fitting in the
 training), which nevertheless cannot be corrected at low rates, an artifact we
 term ``basis break-out".

It is not obvious how to improve performance.
Better training and more training data could improve the quality of the
 predictors we obtain, but we tried many different variations of our training
 methods, and suspect the potential upside here is limited.
The approximations in TF are certainly one potential culprit.
Although TF itself is orthonormal, our lapped transforms are biorthogonal, and
 each basis function has a different magnitude, depending on which post-filters
 are applied.
With the full lapping scheme, it is difficult to even compute the magnitudes of
 the basis functions, as the filter order makes them depend on the size of all
 of ones neighbors (see Figure~\ref{fig:basis-weirdness}).
That makes it impractical to correct for the magnitudes of basis functions
 obtained via TF compared to those obtained with a fixed block size, much less
 for inaccuracies in the shape of the basis functions.
That limits the effectiveness of the predictors we can train.
With the fixed lapping scheme, these issues may be worth revisiting, as might
 the approach of de Oliveira and Pesquet-Popescu.

Instead, we opt to use two much simpler tools.
The first, called Haar DC~\cite{VTEDCMB16}, simply applies our $2\times 2$
 WHT to recursively merge just the DC coefficients from each block up to the
 superblock level.
At the superblock level, DC is predicted using a simple, static linear
 predictor.
This ensures that we retain fine quantization of DC over large regions, since
 the magnitude of the DC basis function doubles with each application of the
 WHT, but the quantizer resolution does not.
It also greatly reduces the number of bits spent on DC.

The second simply copies the first row and/or column of AC coefficients from
 its neighbors, much like MPEG4 Part~2, with the important distinction that we
 use the copied coefficients as predictors \textit{before} quantizing the AC
 coefficients in the current block, not afterwards.
This copying is only done when the neighbors use the same transform block size.
We experimented with using TF to generate predictors when the transform sizes
 differed, but this did not produce any appreciable gains.

Since we already partition each block into multiple frequency bands, and signal
 whether or not to use the prediction in each one (using ``noref" flags, see
 Section~\ref{sec:pvq}), this approach requires no additional signaling, and
 no additional searching.
For the lowest band, which contains portions of both the first row and the
 first column of AC coefficients, we only copy the one that has higher energy.
On typical images, this prediction only reduces rate by $1$ to $2\%$, but on a
 synthetically generated image of a large checkerboard, it reduces rate by
 around $50\%$.
That is, its main utility is to prevent embarrassingly bad performance on
 trivial images.

Together, these two tools are about as effective as the much more complex
 frequency-domain intra prediction schemes of the previous sections.
In the context of the Alliance for Open Media, frequency-domain intra
 prediction is mostly unnecessary, since without lapped transforms, spatial
 intra prediction can be used instead.
That means neither Haar DC nor the simple AC coefficient copying are being
 pursued for integration.
However, the pattern predictor illustrated in Figure~\ref{fig:end-of-show} may
 have some potential as an additional prediction mode.
The difficulty will be justifying the additional line buffers it would require
 in a hardware implementation.

\subsection{Overlapped Block Motion Compensation}
\label{sec:obmc}

There are multiple ways to incorporate motion compensation into a codec with
 lapped transforms.
Tran and Tu originally proposed applying the pre-filter to the reference frame,
 at offsets determined by a motion vector (MV), to produce lapped reference
 pixels~\cite{TT01}.
However, this requires redundant filtering of edges when neighboring MVs do not
 match, especially during the encoder's motion search.
It also directly ties the prediction block size to the transform block size, a
 tie that the most recent video codecs, such as Thor~\cite{FBMDZ16} and
 VP10~\cite{MSBCHLX15}, have reduced or eliminated.

Instead, we produce a complete ``motion-compensated reference frame",
 and then apply exactly the same lapping to this frame as we apply to the input
 image to produce our predictors.
The prediction stage is completely decoupled from the transform stage.
With this structure, we can use any method we wish to create the
 motion-compensated reference frame.
We choose to use a variable-block-sized Overlapped Block Motion Compensation
 (OBMC) scheme~\cite{Ter15a}.
In addition to eliminating blocking artifacts, it also produces a relatively
 uniform prediction error, which, together with our vastly simplified intra
 prediction, mostly eliminates the need for tools such as adaptive transform
 type selection~\cite{HXM13}.
We view this as an advantage, as having fewer alternatives allows for less
 searching at the encoder and a simpler implementation.

The major drawback of this approach is that the motion estimate stage has no
 way to know when the transform stage will choose to discard the prediction
 (using ``noref" flags, see Section~\ref{sec:pvq}).
This is the closest we have to an intra mode in inter frames, as we restrict
 the AC copying scheme described in Section~\ref{sec:simpler-intra} to intra
 frames for simplicity.
In regions that are difficult to predict, it is possible we will waste bits
 producing a prediction that is not good enough for the transform stage to use.

We have considered a number of ideas for mitigating this problem.
We could include using a heuristic to approximate the transform stage's
 decision, an approach commonly used for mode decision in other codecs.
Or, we could use a special reference frame that produces a constant color as
 its prediction, which would then be easy to blend with the predictions of
 other reference frames in the OBMC scheme.
However, we have not implemented these ideas, and do not know how effective
 they would be.

However, one advantage this structure brings is the ability to easily swap out
 one prediction scheme for another.
To compare the objective performance of our OBMC scheme with a traditional
 Block Matching Algorithm (BMA) implementation, we do just that, using the
 open-source Thor codec.
We simply run the entire Thor encoder with residual coding disabled in place of
 OBMC, despite the obvious issues with blocking artifacts (although Thor has a
 loop filter, it only runs when residuals are actually coded).
To make the comparison more fair, we disable Thor's intra prediction modes and
 $64\times 64$ prediction block sizes (which the version of Daala we tested did
 not yet support), and stop coding the relevant bits that would have signaled
 them\footnote{See commit 41f301b1b142 in the \texttt{exp-thor-mc1} branch of
 Timothy B. Terriberry's personal Daala repository:
 \url{https://git.xiph.org/users/tterribe/daala.git}.}.
The result was an increase in bitrate of $7.2\%$ to $9.2\%$, depending on the
 metric.
However, given the complexity of our OBMC scheme, this was not as large as we
 were expecting.
For comparison, at the time we conducted these experiments, Thor spent 
 approximately $10\%$ of its encoding time on motion estimation, while Daala
 spent approximately $70\%$.

We then add back in Thor's intra
 modes\footnote{See commit 35766e08483d in the same branch.}.
This reduces the rate of using Thor's MC by $1.6\%$ to $2.0\%$ (depending on
 the metric, FastSSIM actually increased by $0.5\%$), but it is still worse
 than Daala's.
Although this does not completely eliminate the potential to waste bits on
 predictions that are not used, it gives some idea of the potential
 effectiveness of mitigation strategies in the motion estimation stage.

While these results are interesting, it is not straightforward to perform the
 reverse experiment---putting Daala's OBMC scheme into an encoder that tightly
 couples its search for both optimal predictions and the optimal transform
 coding of the residual from those predictions.
Nor is it clear how to incorporate the usual spatial intra prediction modes
 into our OBMC scheme without violating our structural guarantees against
 blocking artifacts (which would defeat the purpose of the scheme).
Thus, given how much of the encoder we would have to rewrite and how many
 coding tools we would have to discard, we are not currently planning to
 integrate our OBMC scheme into AV1, despite its promise.

\subsection{Perceptual Vector Quantization}
\label{sec:pvq}

Perceptual Vector Quantization~\cite{VT15} is a gain-shape quantization
 technique that also has its origins in the Opus audio codec~\cite{VMTV13}.
Its primary advantage over scalar quantization is that it offers a lot more
 flexibility in automatically controlling the allocation of bits.
This is due to the fact that it extracts several meaningful parameters from the
 vector of transform coefficients: both a \textit{gain}, representing the
 amount of contrast in the signal, and $\theta$, which represents the angle
 between the input vector and a prediction, and thus encodes how well the
 signal is predicted.
PVQ then uses algebraic vector quantization to encode the remaining information
 in the coefficient vector without any redundant degrees of freedom.

Having a meaningful parameter like the gain allows us to implement things like
 activity masking without sending any side information.
The gain directly tells us how much contrast was in the original signal, which
 gives an indication of how visible quantization artifacts will be.
We even split the coefficients into multiple frequency bands, with a separate
 gain and $\theta$ encoded for each one, which allows finer control of this
 effect.
Having separate bands also allows us to independently signal for each one
 whether or not to use the predictor, what PVQ terms a ``noref" flag for ``no
 reference".
Together with $\theta$, this gives a great deal of flexibility in how the
 predictor is used, particularly when the reliability of the predictor varies
 by frequency.

\setcounter{footnote}{0}

Unlike some of the other tools developed for Daala, conceptually PVQ is a
 fairly straightforward drop-in replacement for scalar quantization.
In our original published results on PVQ~\cite{VT15}, we compare it to a
 fairly unsophisticated implementation of scalar quantization (i.e., one that
 we wrote ourselves), with encouraging results.
In AV1, we are starting from the scalar quantization implementation in VP9,
 which has matured and been tuned over the course of many years.

We have begun integrating PVQ into AV1 as an optional
 experiment\footnote{Currently in the \texttt{av1\_pvq} branch of Yushin Cho's
 personal AOM git repository: \url{https://github.com/ycho/aom/tree/av1_pvq}}.
While this integration process touches many parts of the encoder, and is still
 an ongoing process, with many unknown bugs to find and unexpected interactions
 to tune, we have already achieved encouraging results.
Currently, this integration targets PSNR, using flat quantization matrices and
 disabling activity masking.
Only once we are confident this works in all configurations will we attempt to
 enable more advanced features.
We also use a monochrome version of the \texttt{objective-1-fast} test
 sequences, \texttt{objective-1-smc}, since we have not yet properly tuned
 chroma with PVQ in AV1.

With this configuration, on still images, we show gains from $1.3\%$ to
 $2.0\%$, with the largest gains in PSNR\footnote{On \texttt{subset1} using
 commit 1ca568b15e4f for PVQ and commit 636b94545d51 for AV1, which enables
 trellis quantization for intra frames for a fairer comparison.
 We use the High + Latency CQP settings from the testing draft\cite{DNB16}
  with the extra options \texttt{-{}-cpu-used=1} in both cases}.
For video, we currently see $0.65\%$ and $0.55\%$ gains in PSNR and PSNR-HVS,
 but a loss of $0.92\%$ and $0.02\%$ in SSIM and MS-SSIM,
 respectively\footnote{On \texttt{objective-1-smc} using commit 1ca568b15e4f
 for PVQ and commit 636b94545d51 for AV1 again, with the High
 Latency CQP settings from the testing draft\cite{DNB16} with the extra options
 \texttt{-{}-cpu-used=1 -{}-passes=1 -{}-frame-parallel=1} in both cases.}.
While extremely preliminary, these results show that even with minimal tuning,
 and without optimizing for the perceptual metrics for which it was designed,
 PVQ is competitive with scalar quantization.
One potential drawback is that PVQ requires an extra forward transform to
 transform its predictor to the frequency domain.
This was a cost we had already paid in Daala when using lapped transforms with
 frequency domain prediction, but is one we will have to justify in AV1.

\subsection{Chroma from Luma}
\label{sec:cfl}

Chroma from Luma (CfL) has proved to be an effective tool to exploit local
 correlation between the color planes by building a linear model between the
 luma plane and each of the chroma planes in a given block~\cite{EV15}.
Frequency-domain CfL works particularly well in conjunction with PVQ: it allows
 separating off the DC coefficient (equivalent to the offset of the linear
 model) and the scale (equivalent to PVQ's gain parameter).
Instead of trying to build a linear model implicitly without signaling, we find
 that simply using the luma coefficients directly as a shape predictor and
 signaling DC and gain parameters via normal means works even better.
This reduced bitrate of the overall file by $1.4\%$ for equal PSNR in the Cb
 plane and $0.4\%$ for Cr\footnote{See commit 0857e3029854 in the Daala git
 repository.
Our other metrics are only designed to work on luma, which was only marginally
 affected by the change.}.
This is an example of how the flexibility of PVQ simplifies other coding tools.

The other advantage of operating in the frequency domain is that we can compute
 the predictor before performing final reconstruction of the pixels in the
 spatial domain.
That reduces pipeline dependencies in a hardware implementation.
For these reasons, we plan to investigate the integration of frequency-domain
 CfL into AV1 as soon as PVQ is working well.

\subsection{Directional Deringing}
\label{sec:deringing}

Daala has a directional deringing filter that operates as an in-loop filter on
 the final reconstructed output of the encoder~\cite{Val16}.
That makes it very straightforward to adapt to any other hybrid video codec,
 including AV1.
We have already integrated it into AV1 as an experiment
 (\texttt{-{}-enable-dering}) alongside the Constrained Low-Pass Filter
 (CLPF) from Thor (\texttt{-{}-enable-clpf}).
We tested both Daala's deringing filter and the CLPF, as well as applying both
 filters.
For the latter, one filter runs and feeds its output into the other, in
 pipelined fashion.
The results are summarized in Table~\ref{tab:dering-results}, using both the
 low-latency and high-latency configurations of NETVC's proposed testing
 procedures~\cite{DNB16}.

\begin{table}[tb]
\begin{center}
\begin{tabular}{l@{\hspace{0.05in}}rrrrc@{\hspace{0.15in}}rrrr}\toprule
              & \multicolumn{4}{l}{Low-latency}               & & \multicolumn{4}{l}{High-latency}              \\\cmidrule{2-5}\cmidrule{7-10}
Method        & PSNR      & PSR-HVS   & SSIM      &  MS-SSIM  & & PSNR      & PSR-HVS   & SSIM      & MS-SSIM   \\\midrule
CLPF          & $-2.80\%$ & $-1.85\%$ & $-2.77\%$ & $-2.11\%$ & & $-0.97\%$ & $-0.33\%$ & $-0.83\%$ & $-0.48\%$ \\
DD            & $-3.54\%$ & $-2.95\%$ & $-3.95\%$ & $-3.15\%$ & & $-1.59\%$ & $-0.95\%$ & $-1.74\%$ & $-1.17\%$ \\
CLPF, then DD & $-3.87\%$ & $-2.75\%$ & $-3.97\%$ & $-3.03\%$ & & $-1.53\%$ & $-0.55\%$ & $-1.53\%$ & $-0.85\%$ \\
DD, then CLPF & $-4.01\%$ & $-2.96\%$ & $-4.16\%$ & $-3.26\%$ & & $-1.67\%$ & $-0.73\%$ & $-1.70\%$ & $-1.03\%$ \\
\bottomrule\end{tabular}
\end{center}
\caption{Bitrate reductions for the Daala directional deringing filter (DD) and
 Thor's CLPF in AV1, in various configurations.
All numbers are percent reductions in bitrate at equivalent quality for the
 given metric (more negative is better).}
\label{tab:dering-results}
\end{table}

The larger gains for the low-latency configuration are expected, as the use of
 alt-refs in the high-latency configuration has a noise-removal and low-passing
 effect similar to that of the filters.
We expect these gains will improve further, as the side information encoded
 for each superblock (to signal application of the filter and filter strength)
 is not currently entropy coded.
When running both filters in sequence, each one currently makes decisions about
 whether or not to apply the filter and at what strength independently, while a
 joint decision might be better.
Overall, these gains are smaller than when we tested the same filters in Daala,
 most likely because AV1 has less ringing to remove.
Our general impression, however, is that the subjective improvements are still
 larger than the metrics would suggest.

Combining Daala's deringing with Thor's CLPF demonstrates that some gains are
 possible by using both, mostly in the low-latency configuration, but the
 current approach was done for simplicity, and will likely change before AV1 is
 done.
Switching between the filters on a superblock basis instead of pipelining them
 would reduce the amount of line buffers required in hardware, but since they
 do not always operate on the same superblock size, this is not something we
 have attempted.

\subsection{Non-Binary Adaptive Entropy Coding}
\label{sec:entropy-coding}

Daala uses an adaptive entropy coder that, unlike most video codecs, is not
 restricted to coding symbols from binary alphabets~\cite{Ter15b}.
Entropy coding is an inherently serial process, and can be the limiting factor
 in lowering the clockrate required by a hardware implementation (and thus the
 power required).
Every symbol clocked through the entropy coder incurs per-symbol overheads.
The primary motivation of coding symbols from alphabets with more than two
 possible values is to reduce the number of symbols coded to improve hardware
 efficiency.
By reducing the number of symbols we code, we can reduce these overheads as
 well as the number of serial dependencies.

\subsubsection{Non-Binary Coding}

We support alphabet sizes up to $16$ using $15$-bit probabilities.
Larger alphabets are easy to support in software, but imposing a limit makes
 hardware simpler, and also allows SIMD implementations of probability updates.
Thus each symbol coded with our non-binary coder can take the place of up to
 four binary symbols.
This effectively gives us a (small) degree of parallelism.

Although we do not yet have a hardware implementation to use for comparisons,
 we created a small benchmark for testing software entropy coder
 implementations in isolation\footnote{See the \texttt{ectest} branch of Nathan
 E. Egge's personal AOM git
 repository: \url{https://github.com/negge/aom/tree/ectest}}.
Like VP9, AV1 codes symbols using binary trees, with a binary probability at
 each node of the tree.
This benchmark randomly generates $100,000,000$ symbols according to the
 default probability distribution of the tree used for $16\times 16$ intra
 modes.
It then converts the tree into a flat $15$-bit cumulative distribution for use
 with our entropy coder and a flat $10$-bit cumulative distribution for use
 with rANS~\cite{DTGD15}, which is also under consideration for inclusion in
 AV1, and encodes and decodes each symbol with all three methods.
The alphabet size is $10$ and the average number of binary symbols required to
 encode a value from this distribution is $2.71$, with $51.5\%$ of symbols
 requiring only a single binary symbol to code.
The resulting throughput of the encoder and decoder in each method is given in
 Table~\ref{tab:entropy-throughput}.

\begin{table}[tb]
\begin{center}
\begin{tabular}{l@{\hspace{0.15in}}rrr}\toprule
Method   & VP9          & Daala        & rANS        \\\midrule
Encoding & $107.2$ Mbps & $192.2$ Mbps & $366.8$ Mbps \\
Decoding & $112.6$ Mbps & $144.9$ Mbps & $181.5$ Mbps \\
\bottomrule\end{tabular}
\end{center}
\caption{Encoder and decoder throughput for several entropy coding methods.
All numbers are in megabits per second (higher is better).}
\label{tab:entropy-throughput}
\end{table}

This shows a significant speed-up for Daala over VP9, and rANS shows an even
 larger speed-up, particularly for encoding.
However, rANS also brings several practical complications, such as the
 requirement of a $33$~kbit lookup table in its current implementation and the
 requirement that symbols be decoded in the opposite order from which they are
 encoded.
The latter requires buffering all of the symbols for a tile in the encoder,
 which is expensive for hardware.
We do not claim that these software implementations of each method are fully
 optimized, nor that this single distribution is representative of the entire
 codec, however we expect this general trend to hold.
The exact speed-ups depend a good deal on the average number of binary symbols
 represented by a single non-binary symbol, where we feel there is significant
 room for improvement in the design of AV1.

\subsubsection{Adaptive Coding}

Currently, like VP9, AV1 uses static probabilities for an entire frame.
These probabilities are initialized to default values, then automatically
 updated using the statistics of the previous frame, and optionally overridden
 by values explicitly coded in the frame header.
In its current implementation, the AV1 encoder performs R-D optimization of the
 entire frame, recording all of its decisions.
Then it uses the statistics of these decisions to compute optimal probabilities
 for that frame, and makes a R-D optimal decision about whether or not to
 encode an explicit override for each one.

This has several drawbacks for interactive applications.
Because such applications have to be robust in the face of packet loss, they
 cannot use the feed-forward updates of the probabilities using the statistics
 from prior frames.
That means all probability updates must be explicitly coded from the defaults
 for every frame.
This adds bitrate overhead on the order of $5.5\%$.
It also means that the encoder must buffer the symbols to code for an entire
 frame, and cannot send out a single packet until the whole frame has finished
 encoding.
As with rANS, this buffering is expensive.
However, unlike rANS, it is possible to modify the implementation to encode
 with estimated probabilities rather than optimal probabilities.
We have not modified the encoder to do so, but expect it would incur a
 similarly large overhead.

Instead, we are investigating using adaptive probabilities.
Currently, Daala adapts most probabilities using simple frequency counts,
 periodically rescaling them as the total exceeds $15$ bits.
Since the total frequency count is generally not a power of two, we adapt the
 method of Stuiver and Moffat~\cite{SM98} to coding these values without any
 multiplications or divisions (except that we over-estimate probabilities for
 symbols at the beginning of the alphabet, instead of the end).
This method is only approximate, but has an overhead of around $1\%$ to $2\%$
 compared to the theoretical entropy, which is similar to that of CABAC.

When the total frequency count is a power of two, we use
 $15\times 16\rightarrow31$ multiplies to code symbols with negligible
 ($< 0.01\%$) overhead.
If we can adapt the probability distributions while leaving their total
 constant, we can avoid the overhead of the non-power-of-two approach.
Let $f_i$ be a cumulative distribution of an alphabet of size $M$ with total
 $T$, such that $0 < f_0 < f_1 < ... < f_i < ... < f_{M-1} = T$.
Then, if we encode or decode symbol $j$, we propose updating the distribution
 $f_i$ with an update rate of $2^r$ using the following rule:
\begin{align}
f_i & = \begin{cases}
f_i-\bigl\lfloor\frac{f_i-i+2^r-2}{2^r}\bigr\rfloor\ , & i < j \\
f_i-\bigl\lfloor\frac{f_i-T+M-1-i}{2^r}\bigr\rfloor\ , & j \le i\ .
\end{cases}
\end{align}
This rule adjusts values of $f_i$ downward when $i < j$ and upward when
 $j \le i$, but ensures that the difference between $f_i$ and $f_{i+1}$ is
 always at least $1$, and maintains the total, $T$.
So long as the difference is larger than $1$, it can continue to adjust the
 corresponding probability, regardless of the rate, $2^r$.
While this update rule looks much more expensive than using simple frequency
 counts, with tables built around the constants $M$, $T$, $2^r$, $i$, and $j$,
 it can be reduced to two vector additions and a vector shift, which is
 comparable to a frequency count update and a rescaling operation.

We implemented this adaptation scheme, with some additional rules to speed up
 the adaptation of the first few symbols coded in each context.
Just converting some of the symbols coded in Daala to use this
 scheme reduced bitrates by $0.3\%$ to $0.4\%$\footnote{See the
 \texttt{dyadic\_adapt8} branch of Jean-Marc Valin's personal Daala git
 repository: \url{https://github.com/jmvalin/daala/tree/exp_dyadic_adapt8}}.
Whether or not the complexity of the more expensive update rule is worth the
 reduction in entropy coder overhead is still an open question.

We think that some of these ideas merit inclusion in AV1, though which exact
 direction the Alliance ultimately goes depends on exact engineering trade-offs
 and implementation costs, especially in hardware.
We will continue to experiment with these ideas and get more feedback from our
 partners in the Alliance.

\subsection{Rate Control}
\label{sec:rate-control}

Daala extends a bit-reservoir rate control system refined from the
 Vorbis audio codec and Theora video codec.
The rate control is conceptually simple, flexible, stable, precise, and adapts
 quickly.

\subsubsection{Rate Control Model}

Daala controls rate by manipulating the quantization parameter, and thus
 quantizers, used in encoding.
A simple model predicts bit usage for a given frame for any quantizer.

\begin{align}
R & = scale \cdot Q^{-\alpha}
\end{align}

$R$ represents the rate of a frame in bits, $Q$ is the actual linear quantizer
 (a function of the quantization parameter), and $\alpha$ is a fixed modeling
 exponent chosen for an entire sequence based on frame type and target rate
 (in bits-per-pixel).
We measure the parameter $scale$ for each frame type during encoding.

\subsubsection{Estimating $scale$}

After encoding a frame, we know $R$ and $Q$, and $\alpha$ is fixed.
As a result we can solve the above relationship for $scale$.

\begin{figure}[htb]
\center
\begin{subfigure}[b]{0.5\textwidth}
\center
\includegraphics[width=0.9\columnwidth]{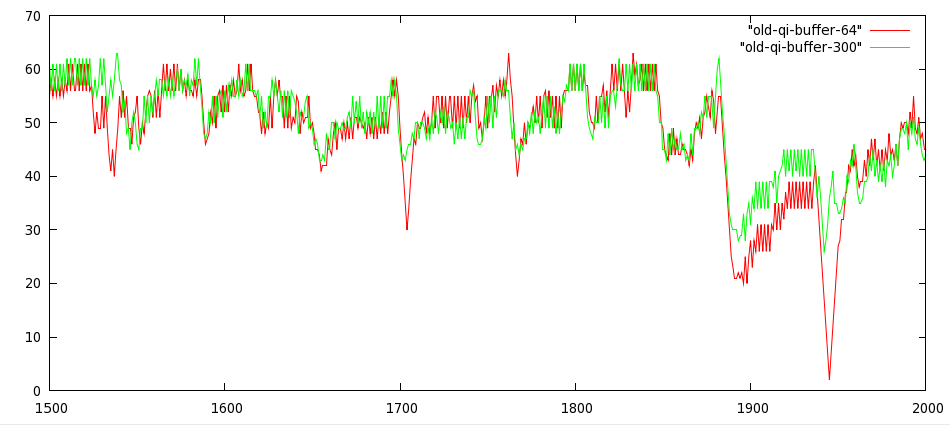}
\caption{Exponential Moving Average}
\label{fig:old-onepass}
\end{subfigure}%
\begin{subfigure}[b]{0.5\textwidth}
\center
\includegraphics[width=0.9\columnwidth]{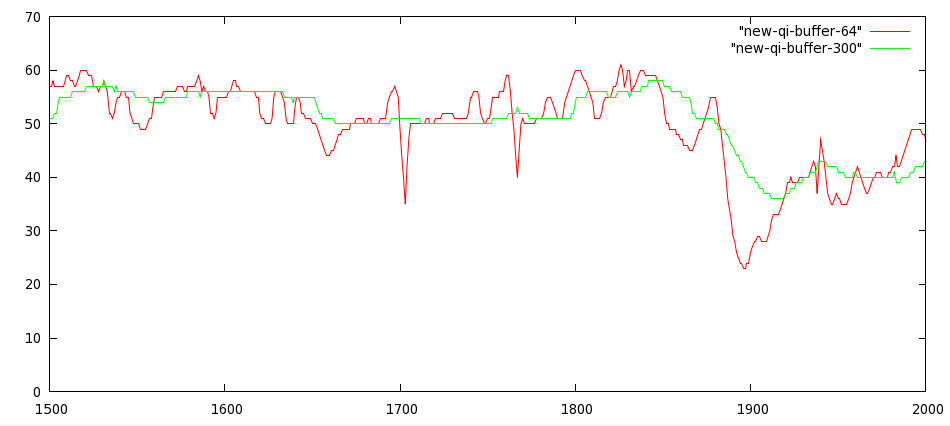}
\caption{$2^\textrm{nd}$ Order Bessel Filter}
\label{fig:new-onepass}
\end{subfigure}%
\caption{(\ref{sub@fig:old-onepass}) Smoothing $scale$ modeling and quantizer
 choice using an exponential moving average produces unstable/oscillatory
 behavior detrimental to objective and visual quality.
(\ref{sub@fig:new-onepass}) A $2^\textrm{nd}$ order Bessel filter damps the
 oscillations while preserving reaction speed.
In addition, the Bessel filter's time constant may be adjusted independent of
 buffer size to produce any desired reaction speed.}
\label{fig:onepass-filter}
\end{figure}

We smooth this measured $scale$ value using a digital approximation of a
 second order Bessel filter.
The filter damps noise and oscillations while preserving reaction
 speed (see Figure~\ref{fig:onepass-filter}), using a time constant
 that allows full-scale reaction in half the buffer interval.
We adjust the reaction time to allow faster adaptation at the beginning of a
 sequence.

Daala models a separate $scale$ parameter for each frame type and so maintains
 a Bessel filter for each of keyframes, long-term reference (golden) frames,
 regular P frames, and B frames.

\subsubsection{Choosing $Q$}

For every frame, Daala constructs a bit usage plan for the entire remaining
 buffer interval by predicting the frame type of each subsequent frame and
 summing the bits used for each of these frames according to our prediction
 model.
The goal of the bit usage plan is to hit a fixed buffer fullness level at the
 last keyframe in the buffer interval.

\begin{align}
\label{eq:rate-plan}
R_{total} & = \sum_{i}N_i \cdot scale_i \cdot Q_i^{-\alpha}
\end{align}

Here, $N_i$ is the number of frames of each type, $scale_i$ are the
 Bessel-smoothed scales corresponding to each frame type, and quantizers
 $Q_i$ are functions of the quantization parameter.
The encoder simply searches quantization parameters until $R_{total}$ hits the
 desired bit usage target.

After encoding each frame, we discard the plan and construct a new plan
 starting from the next frame.

\subsubsection{Two-Pass Rate Control}

Most modern video encoders offer two-pass rate control which makes a first pass
 over a video file to collect metrics, and performs encoding in a second pass
 using knowledge of the complete video content.
Daala's two-pass rate control is a simple modification of the (one-pass,
 causal) rate control described thus far.

Rather than smoothing $scale$ values for each frame in a Bessel filter, we
 simply measure and save the $scale$ values during the first pass, and sum
 these empirical measured values during second pass encoding to use in place of
 the product $N_i\cdot scale_i$ in equation~\ref{eq:rate-plan}.
We also add a fixed offset to correct for consistent over/under estimates.
Everything else is just as in one-pass.

\subsubsection{Chunked Two-Pass}

Video hosting services and sites such as YouTube typically encode files by
 splitting them into many small chunks of one to five seconds each, and then
 encoding each chunk independently in parallel to reduce encoding time.

Daala's rate control model allows the collection of complete file metrics from
 each chunk in a first pass, allowing the second pass encoding of each chunk to
 take into account the entire sequence.
This produces a chunked two-pass rate control with quality performance similar
 or identical to a monolithic two-pass encoding.

We plan to investigate integration of this rate control method as an
 alternative to AV1's existing rate control.
We do not yet know which approach will prove more effective, but conceptually
 Daala's is simple and easily adaptable to a broad array of use cases.

\section{CONCLUSION}

\begin{figure}[htb]
\center
\begin{subfigure}[b]{0.75\textwidth}
\center
\includegraphics[width=\columnwidth]{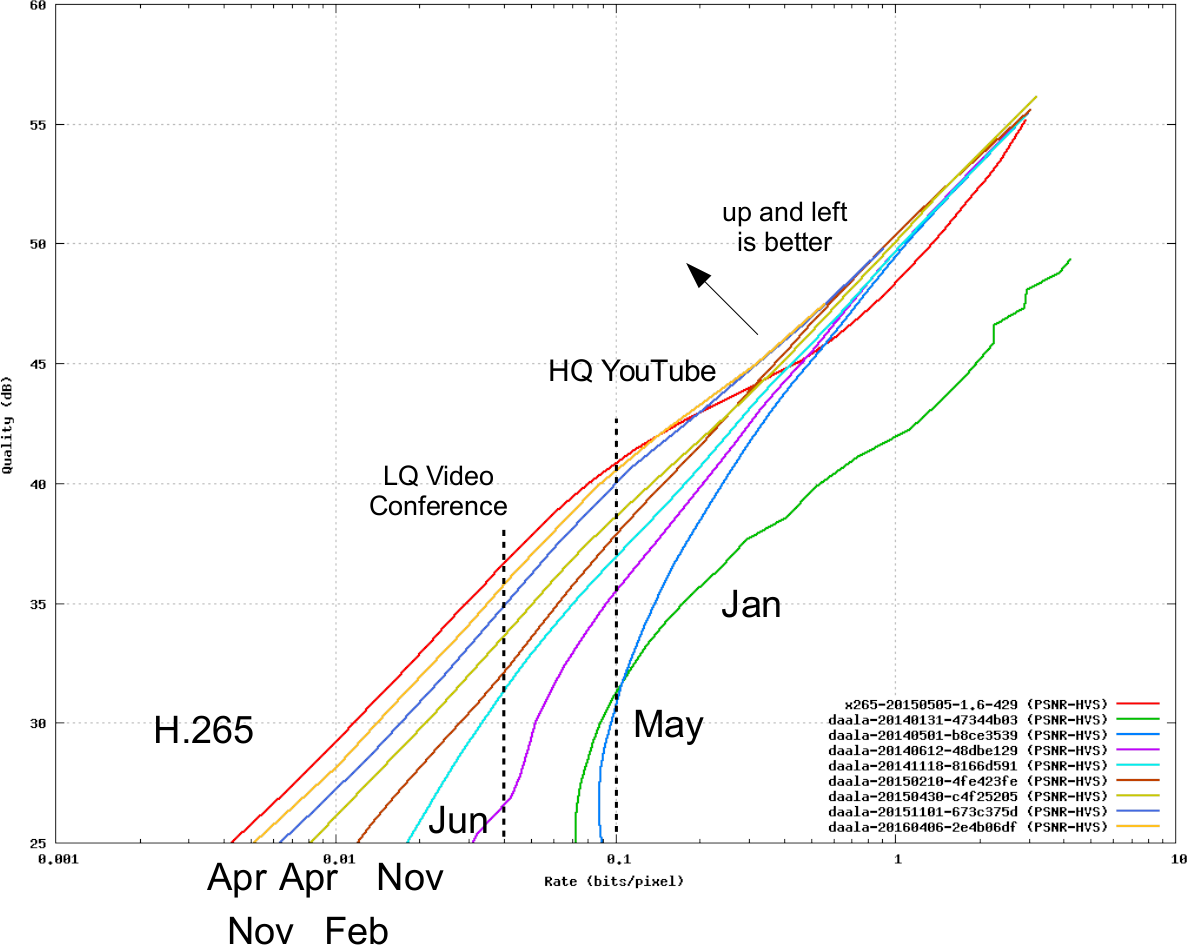}
\caption{PSNR-HVS-M}
\end{subfigure}
\begin{subfigure}[b]{0.75\textwidth}
\center
\includegraphics[width=\columnwidth]{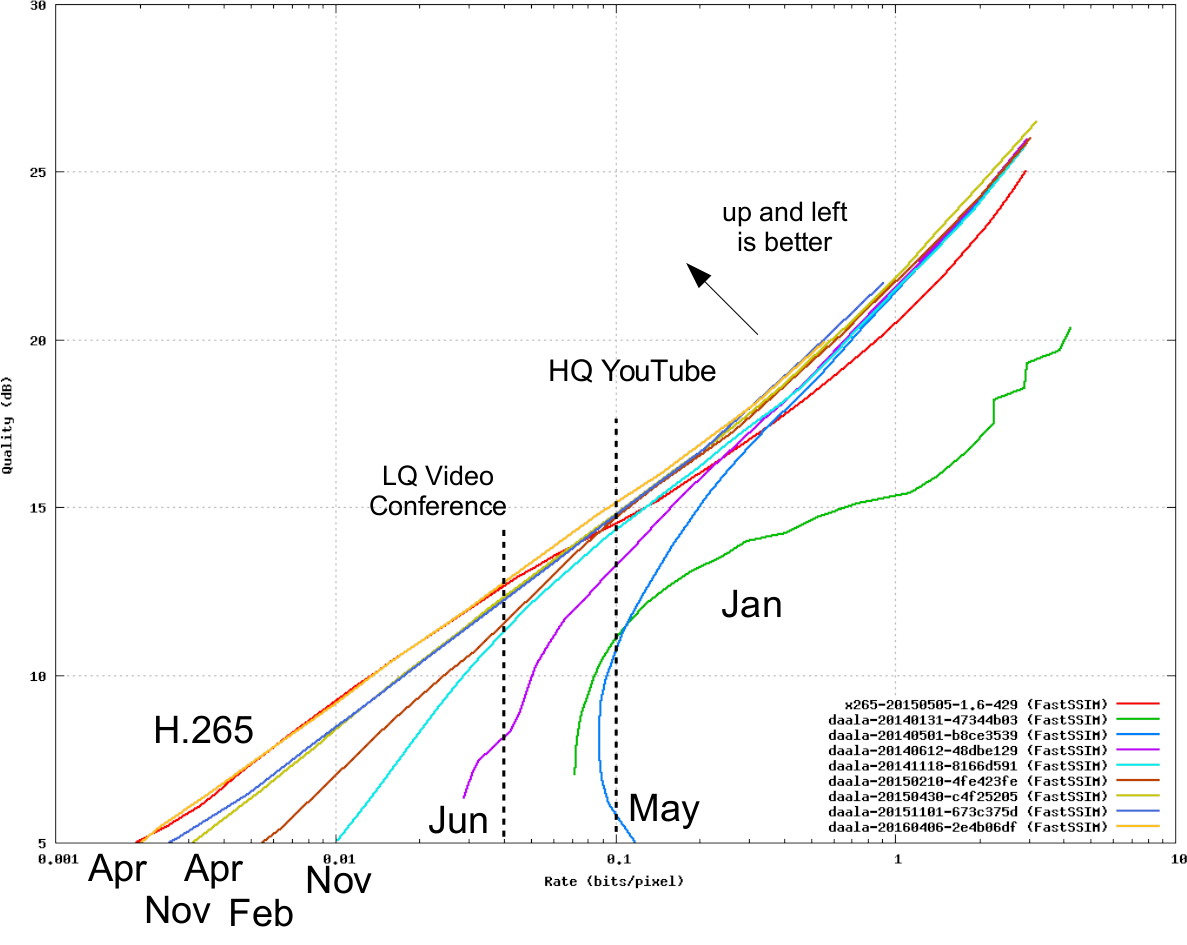}
\caption{Multiscale FastSSIM}
\end{subfigure}%
\caption{Progress of the Daala project vs. x265 from Jaunary 2014 to April
 2016.}
\label{fig:daala-progress}
\end{figure}

Overall, the Daala project has produced many important lessons and many
 promising technologies.
Given that most of these are new or relatively unproven, we consider the
 results we were able to obtain with them encouraging.
These results have steadily improved for the past several years (see
 Figure~\ref{fig:daala-progress}), and the overall Daala design may ultimately
 prove viable in the longer term.
Several of the individual tools are also under active consideration for
 inclusion in AV1, the Alliance for Open Media's first video codec.
We will continue to improve them as development of AV1 proceeds, with the hope
 of creating a truly state-of-the-art royalty-free codec.

\section*{ACKNOWLEDGMENTS}

The authors would like to thank Steinar Midtskogen for his assistance in
 generating the deringing and CLPF results, as well as James Bankoski, Alex
 Converse, and Yaowu Xu for their assistance integrating code into AV1, in
 addition to all of the other volunteers that contributed to the open-source
 Daala project.

\appendix

\clearpage

\bibliography{daala-spie-adip2016}

\end{document}